\documentclass[onecolumn,12pt]{IEEEtran}
\ifCLASSINFOpdf
	\usepackage[pdftex]{graphicx}
	\usepackage{epstopdf}
	\graphicspath{{./figures/}}
	\DeclareGraphicsExtensions{.pdf,.jpeg,.png,.eps}
\else
	\usepackage[dvips]{graphicx}
	\graphicspath{{./figures/}}
	\DeclareGraphicsExtensions{.eps}
\fi

\usepackage[caption=false,font=footnotesize]{subfig}
\usepackage{dblfloatfix}

\usepackage{url}
\title{Desktop and Mobile Web Page Comparison: Characteristics, Trends, and Implications
	\thanks{T. Johnson and P.~Seeling are with the Dept.~of Computer Science,
		Central Michigan University, Mount Pleasant, MI 48859, Email: \texttt{johns4ta@cmich.edu,pseeling@ieee.org}}
}

\author{Troy Johnson and Patrick~Seeling
	\thanks{Please direct correspondence to P.~Seeling.}
	\thanks{Supported in part by an Early Career Grant from the Office of Research and Sponsored Programs at Central Michigan University.}
	\thanks{\textcopyright 20xx IEEE. Personal use of this material is permitted. Permission from IEEE must be obtained for all other uses, in any current or future media, including reprinting/republishing this material for advertising or promotional purposes, creating new collective works, for resale or redistribution to servers or lists, or reuse of any copyrighted component of this work in other works.}
}

\begin{document}
\maketitle

\begin{abstract}\boldmath
The broad proliferation of mobile devices in recent years has drastically changed the means of accessing the World Wide Web.
Describing a shift away from the desktop computer era for content consumption, predictions indicate that the main access of web-based content will come from mobile devices.
Concurrently, the manner of content presentation has changed as well; web artifacts are allowing for richer media and higher levels of user interaction which is enabled through increasing access networks speeds.

This article provides an overview of more than two years of high level web page characteristics by comparing the desktop and mobile client versions.
Our study is the first long-term evaluation of differences as seen by desktop and mobile web browser clients.
We showcase the main differentiating factors with respect to the number of web page object requests, their sizes, relationships, and web page object caching.
We additionally highlight long-term trends and discuss their future implications.

\end{abstract}


\section{Introduction} 
\label{s:intro}

The broad proliferation of web-based services and the trend to outsource formerly server-based and/or locally maintained services into ``the cloud'' have significantly altered the manner of web page designs and compositions.
As the nature that underlies a typical web page changed over time, so have their characteristics and resulting implications for the networks that transport them.
In turn, a plethora of popular applications and interactions that are performed by users are enabled using web-based services and have fueled the growth of Internet traffic.

Popular past web characteristics, model usages, and performance analysis approaches, see, e.g., \cite{BaCr98} or, more recently, \cite{LiZhZhChGr10}, do hence not necessarily reflect the current state of the World Wide Web anymore.
While Cisco, Inc. predicts that video will account for the largest portion of networked traffic (and its growth) in the near future~\cite{Ci13}, consumer-based web traffic is forecast to exponentially increase during the same time frame as well, putting an overall burden on access networks.

Longitudinal studies have recently emerged that target capturing the dynamic behavior of the World Wide Web over time, such as \cite{CaAlPa10}.
In~\cite{IhPa11}, the authors investigate five years (2006-2010) of fixed web site traffic captured through a proxy system and found significant impacts as a result of interactions.
Furthermore, they note that the overall loading time of web pages has been reduced due to higher levels of caching and an increase in concurrent connections made by desktop browsers.
Similarly, the complexity of web sites was evaluated in~\cite{BuMaSe11,BuMaSe13}. The authors found that the number of objects that were loaded (independent from their relative location) has the highest impact on web page load times.
Interestingly, the authors additionally found that for their dataset, ($i$) non-landing pages tended to be less complex and ($ii$) mobile web pages tended to be of lesser complexity.

The broad proliferation of mobile devices that can connect to Internet-based services has given rise to considerable amounts of data that are exchanged with web-based services. 
Several predictions indicate that there will be a continuous increase in the demand for mobile data, see, e.g. \cite{Ci13}, and that access to web-based services will soon be mainly performed through mobile devices.
While the outlined recent studies and ongoing works mainly investigate the traditional desktop-based web access, the emergence of mobile web access gives rise to a new set of problems that are direct derivatives of the characteristics of mobile devices.

In a recent overview of the battery impact that different web page elements have on the power consumption of mobile  devices, the different elements of web sites were found to contribute differently to the power consumption, see~\cite{ThAgNiBoSi12}.
For a set of popular web pages, an energy profiling for the different elements of web pages was performed that identified Java Script and CSS as the main culprits for the power consumption when browsers are rendering mobile web pages.
The composition in addition to the number of objects and bytes overall hence have a direct relation to the web-related performance of mobile devices.

Our contribution fills the currently existing gap of an in-depth evaluation of the trends that emerge for desktop and mobile client versions of web pages, for which we cover the time period from mid-2011 to mid-2013.
We demonstrate the similarities and disparities of the current fixed and mobile web and outline overall notable characteristics and trends that web page developers as well as networking researchers and practitioners should consider in their respective optimization efforts.

The remainder of this paper is organized as follows. 
In the next section, we describe the underlying dataset from httparchive.org and how we processed it.
Subsequently, we compare the characteristics for desktop and mobile clients with respect to web page objects, sizes, and caching in Section~\ref{s:compare}.
We discuss the results and implications in Section~\ref{s:discuss} before we conclude in Section~\ref{s:conc}.

\section{Data Set and Pre-Processing}
\label{s:dataset}
\begin{figure}
	\centering
	\includegraphics[width=.45\linewidth]{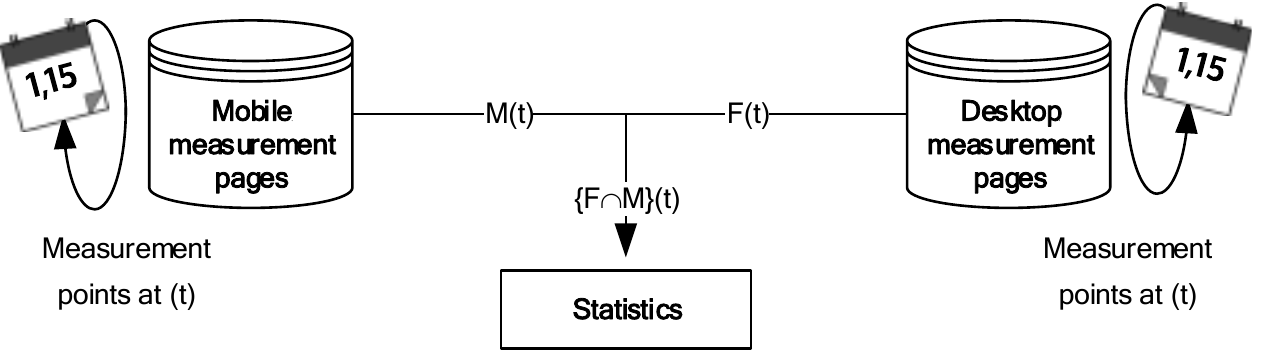}
	\caption{Approach for gathering the general statistics evaluated in this contribution: based on the original archives, we compare the fixed and mobile web pages contained in each bi-weekly measurement and evaluate only those that can be found in desktop and mobile web page requests to allow for a direct comparison.}
	\label{fig:setup}
\end{figure}

In this paper, we utilize the \url{httparchive.org}~\cite{ht13} publicly available dataset of captured web performance metrics. 
As an industry-supported project, its goal is to provide ``a permanent repository of web performance information such as size of pages, failed requests, and technologies utilized.''
The overall starting points are the initial client view statistics, i.e., non-cached web page views, that are gathered by the httparchive.org projects at the beginning and in the middle of each month.
We illustrate the overall process in Figure~\ref{fig:setup}.
As with any project that has to evolve over time to account for changes in technologies, some underlying measurement setups have to change over time as well.
In the remainder of this section, we describe the initial gathering and processing of this dataset in greater detail.

\subsection{General Notes}
As we utilize the measurements over a significant amount of time from June 1, 2011 to July 15, 2013, we initially note that some of the underlying measurement configurations have changed over time, which includes multiple facets, such as Unique Resource Locators (URLs), browser versions, connection speeds, or incorporation of `lazy loading' of resources.
We reason that overall, however, the changes made were reflecting industry trends as well as personal connectivity trends (such as modified access network speeds) and can be seen as representative of the typical connection scenarios for the World Wide Web.
We refer the interested reader to the online documentation of the \url{httparchive.org} project for more details pertaining to  the measurement setup used and detailed information about changes.

\subsection{Web Page Selection and Processing}
We select the available page statistics for both desktop (fixed) and mobile web pages over the range of more than two years; the available points in time are at the beginning and the middle of each month.
This results in a total of 53 datasets each for fixed and mobile web clients, respectively.
We initially determine all web sites that are common for both archives, i.e., web sites that for each measurement time were evaluated for both, fixed and mobile clients, to allow for a comparative evaluation of their described metrics.
For each of the measurement times, we subsequently aggregate the measured web page characteristics over all web sites that were selected in a particular client role -- as a result, we derive a representative average snapshot of the characteristics that make up ``the web'' as accessed using different web clients (represented by different user agent strings) over time.

\subsection{Time Variability of the Dataset}
\begin{figure}
	\centering
	\includegraphics[width=.5\linewidth]{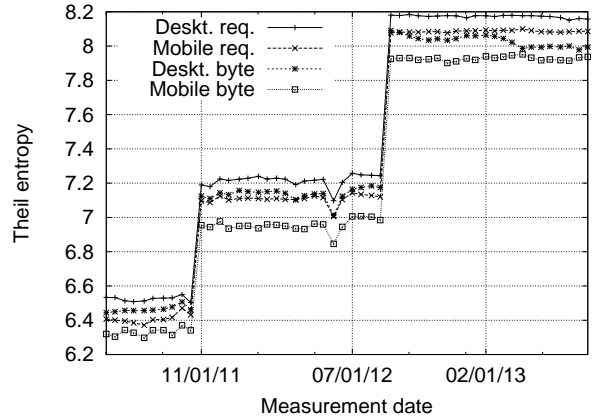}
	\caption{Theil entropy for the number of objects requested and the number of bytes for web pages present in the httparchive.org's desktop and mobile client data files.}
	\label{fig:theil}
\end{figure}

As an initial description of the joint dataset we evaluate in the remainder of this article, we employ the Theil index~\cite{Th72} to determine the equality or inequality of the desktop and mobile client datasets with respect to the diversity of values.
Specifically, we calculate the entropy of the Theil population for the number of web requests and the total number of bytes for each access mode over time, as we are more interested in the overall quality of the selected data subset (i.e., statistics for pages that can be found in both datasets).
We illustrate the result as function of time in Figure~\ref{fig:theil}.
We observe that the four entropy curves are fairly close to each other, with significant ``jumps,'' i.e., changes of their overall level, at two distinct measurement points. 
These ``jumps'' occur at times where the underlying number and source set of the httparchive.org base URLs were changed, namely ($i$) switching to the Alexa top 1000000 web sites in November 2011 and ($ii$) increasing the number of URLs evaluated in September 2012, see~\cite{ht13}.
We note that these changes, however, affect all client modes in a similar manner by significantly increasing the diversity of measurement results, which is immediately visible from Figure~\ref{fig:theil}.
Overall, we note that the Theil entropy levels for the desktop client values are slightly higher than their mobile client counterparts; however, they remain within close range.
This closeness in the calculated Theil entropy additionally motivates us to focus on the overall averages of the values we compare in the remainder of this paper.

\section{Comparison of Desktop and Mobile Web Page Characteristics}
\label{s:compare}
In this section, we compare the average values we obtained from the joint httparchive.org datasets for the desktop and mobile web client versions of the same set of web pages requested, as given by their URLs.

\subsection{Average Number of Web Site Objects}
\label{ss:objects}
\begin{figure}
\centering
	\subfloat[Average total number for desktop and mobile client versions.]{\includegraphics[width=.45\textwidth]{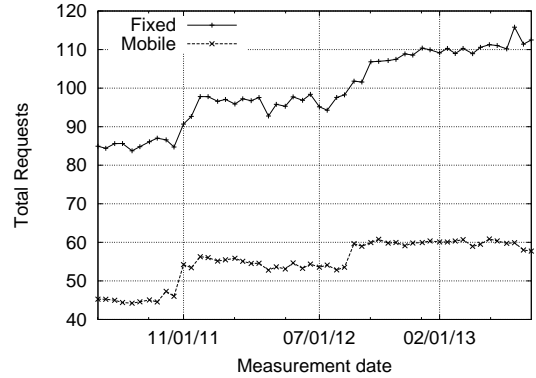}}\\
	\subfloat[Average number by category for desktop client versions.]{\includegraphics[width=.45\textwidth]{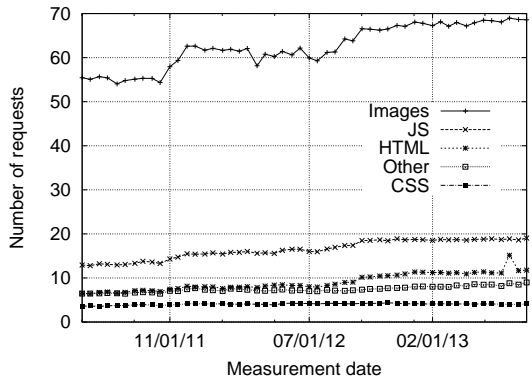}}\qquad
	\subfloat[Average number by category for mobile client versions.]{\includegraphics[width=.45\textwidth]{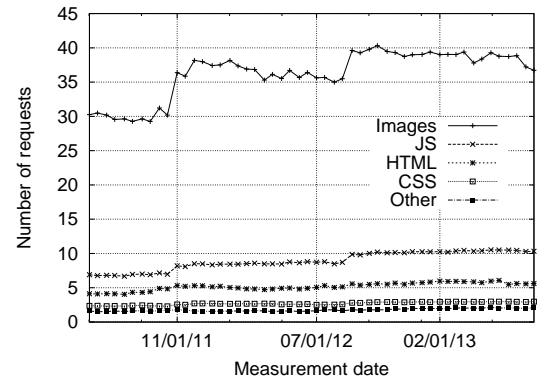}}\\
\caption{Average total number of requests for objects constituting a web page in desktop and mobile versions and decomposition into  HTML, CSS, Java Script, Images, and Other object categories.\label{fig:requests}}
\end{figure}

Initially, we investigate the average number of objects requested when accessing a web page for the first time and illustrate the results in Figure~\ref{fig:requests} for both client types.
We initially note that the desktop versions exhibit a significantly larger number of requested objects in comparison to the mobile counterpart.
We additionally observe an overall continuous increase, which tends to have slight ``jumps'' in certain time frames (more pronounced for mobile client requests) and remains steady to only slowly growing for subsequent time periods.
The slight increases are in line with the earlier observations concerning changes to the underlying web pages constituting the measurement points in time.

Most notably, we witness the increase to over 100 requests on average per web page requested by a desktop client and its continued growth since the second half of 2012.
Within the same time period in 2012, the mobile web request counterpart rose to over 60 requests, with a steadily trend thereafter. 
The implications are manifold, as the continued increase in the average number of web objects requested likely results in additional networking overheads (such as connections setups or DNS resolutions) and could have significant negative power consumption impacts.


Investigating the origins of the average number of web object requests in greater detail, we illustrate the composition of the average number of objects requested separated into the categories of HTML, CSS, Java Script, Images, and Other objects over time in Figure~\ref{fig:requests}~(b) for desktop client requests.
The Other category contains items such as fonts, flash, as well as any remaining downloaded web page components.

We initially observe a rising trend for all categories, with the image category accounting for the most requests and the remaining ``other'' category continuing somewhat steady over time.
Keeping the overall increasing trend in mind, we note the distribution between the different categories of objects remains rather steady over time (with categories in desktop client requests accounting for approximately the following long-term averages: Images 74\%, Java Script 19\%, HTML 11\%, Others 9\% , and CSS 5\%).

For mobile requests, on the other hand, we observe in Figure~\ref{fig:requests}~(c) a nearly identical order at a lower level (with categories in mobile client requests accounting for approximately the following long-term averages: Images 80\%, Java Script 20\%, HTML 11\%, CSS 6\%, and Others 4\%).
We conclude from these numbers that for both scenarios, the main culprit for web requests is provided by the images contained in web sites, and to a lesser extend the oft-mentioned Java Script, found in modern AJAX and HTML5 based pages in fixed as well as mobile environments.

\subsection{Average Web Page Sizes}
\label{ss:bytes}
\begin{figure}
	\centering
	\subfloat[Average total amount for desktop and mobile client versions.]{\includegraphics[width=.45\textwidth]{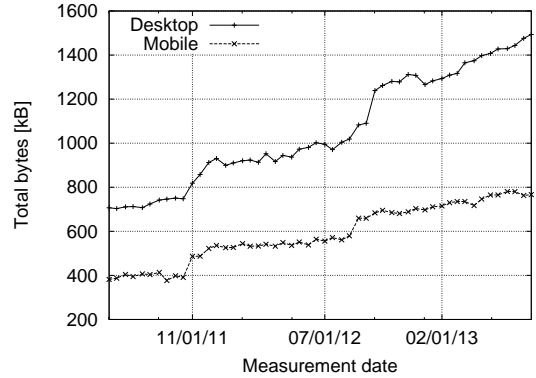}}\\
	\subfloat[Average amount by category for desktop client versions.]{\includegraphics[width=.45\textwidth]{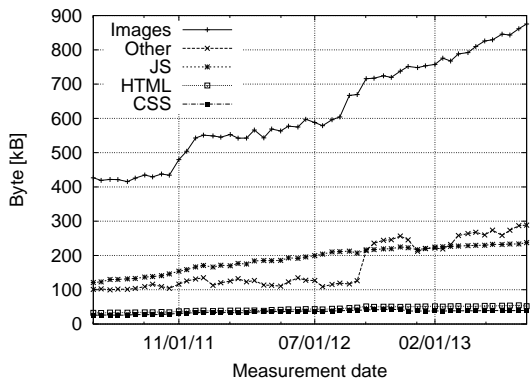}}\qquad
	\subfloat[Average amount by category for mobile client versions.]{\includegraphics[width=.45\textwidth]{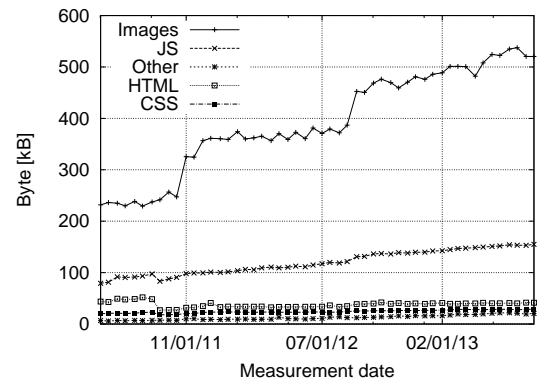}}\\
	\caption{Average total number of bytes constituting a web page in desktop and mobile versions and decomposition into  HTML, CSS, Java Script, Images, and Other object categories.\label{fig:sizes}}
\end{figure}

We now shift our view to the average sizes of desktop and mobile client requested web pages over time as well as their most contributing factors.
We initially illustrate the total number of bytes that were required on average to download a web page to a fixed and a mobile client in Figure~\ref{fig:sizes}.

We observe that the fixed sizes are always larger than their mobile counterparts and that both continuously increase over time. 
This leads to the average web page size to have increased from 0.71 MB in June 2011 to 1.43 MB in June 2013 for desktop client and from 0.38 MB to 0.78 MB for the mobile counterparts in the same time frame.
In greater detail, we note that the average increase in average web page sizes over time for fixed web pages during the evaluated time period is around 3\% per month or approximately 41\% per year, while the mobile counterparts feature an increase of 3\% or 40\%, respectively.
We also find that the average mobile web page sizes currently trail their desktop counterparts by approximately 1.5 years, which likely will increase due to the difference in average growth rates (1.57\% vs. 1.47\% on average in every two week time period for fixed and mobile web pages, respectively).

Investigating the origins of the average number of web page sizes, we illustrate the composition of the average number of bytes requested separated into HTML, CSS, Java Script, Images, and Other object categories over time in Figure~\ref{fig:sizes}~(b) for desktop client requests.
We immediately note that while all components are increasing in size over time, the main contributors to the growth are the Images and Other web page object categories.
More specifically, we note that the three main contributors to total web page sizes are images, other components, and Java Script. 
While the number of bytes contributed by Java Script has slowed in growth, images (which have almost doubled in the number of bytes) and other web page items (which more than doubled) continue their growth over time.
We can attribute this behavior to the richer web experience that users demand, with additional interactivity and visually stimulating appearance.
HTML and CSS components of web pages, on the other hand contribute only a minimum amount of data, with little increase over time.

Shifting the view to the average sizes of the web page components for mobile clients, we also note an overall growth trend for all categories as illustrated in Figure~\ref{fig:sizes}~(c).
For mobile clients, however, the overall average web page sizes stem mainly from images and secondly Java Script. 
The remaining three categories contribute significantly less data and exhibit a slower growth.
It is interesting to note from comparison with the desktop counterparts, that the average mobile client image sizes are approximately two thirds of their desktop counterparts.
Unlike the average number of requests, however, the average number of bytes clearly identifies Java Script for mobile devices as one of the main causes of increased web page sizes.

\subsection{Data per Web Page Object}
\begin{figure}
	\centering
	\subfloat[Average total amount for desktop and mobile client versions.]{\includegraphics[width=.45\textwidth]{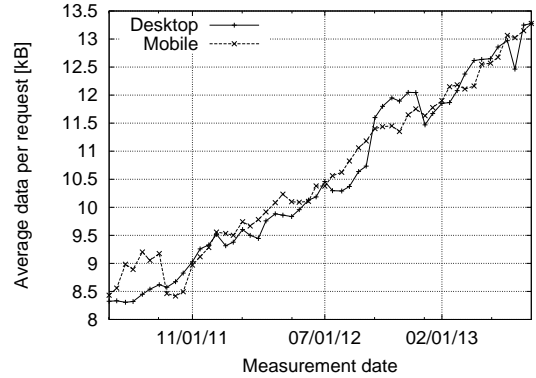}}\\
	\subfloat[Average amount by category for desktop client versions.]{\includegraphics[width=.45\textwidth]{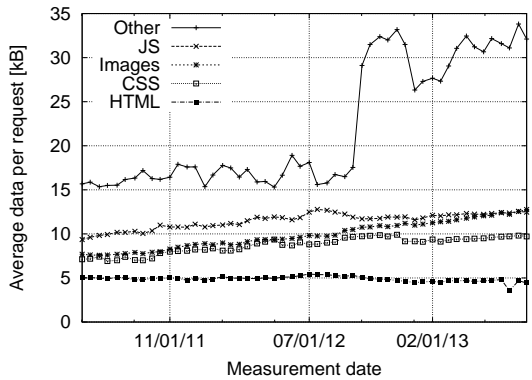}}\qquad
	\subfloat[Average amount by category for mobile client versions.]{\includegraphics[width=.45\textwidth]{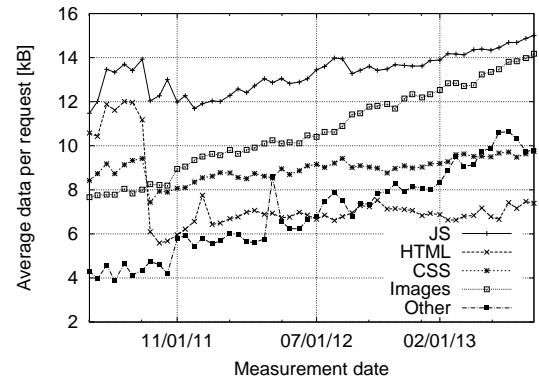}}\\
	\caption{Average number of bytes per web page object request for desktop and mobile versions and decomposition into  HTML, CSS, Java Script, Images, and Other object categories.\label{fig:relative}}
\end{figure}

Combining the two former evaluations, we now evaluate the average web page object request size.
We illustrate the overall and categorized views in Figure~\ref{fig:relative}.
For the average number of bytes per web page request, illustrated in Figure~\ref{fig:relative}~(a), we note a steady increase for both desktop and mobile client requests. 
In greater detail, we note that the overall trend for both request origins is rather linear and close.
In turn, we note that both feature an almost identical overall growth rate of about 0.9\% every two weeks.

Next, we investigate the behavior more closely by evaluating the HTML, CSS, Java Script, Images, and Other categories separately for the desktop and mobile clients in Figures~\ref{fig:relative}~(b) and \ref{fig:relative}~(c), respectively.
For desktop requests, we note that the highest number of bytes per request can be observed for the Other category, with a pronounced ``jump'' in the time frame of base URL adjustments in the underlying dataset. 
In turn, we reason that more components, such as Flash or font data, are included in the larger underlying set of web pages evaluated, which results in the current level of around 33 kB per request in this category.
For Java Script and CSS data per request, we observe an initially growing trend, which has slowed over time and now is around 13 and 10 kB, respectively.
Images, on the other hand, continue their growing trend and now feature about 13 kB per request.
The actual HTML markup request size has remained relatively constant at around 5 kB.

For requests from mobile clients, on the other hand, we note that the largest amount per request is observed for Java Script, with currently growing larger sizes around 15 kB.
The Images and Other categories also feature a significant growth for mobile client requests, which currently are around 14 and 10 kB, respectively.
The HTML and CSS categories are fairly stable with only minimal growth over time, which is similar to the trends observed for the desktop clients.

In comparison, we note that while both average client request sizes are comparable with respect to their level and growth rates, the actual composition is significantly different. 
While desktop request sizes are mainly determined by the Other category data, such as Flash or fonts, and increasingly images, mobile requests are mainly characterized by Java Script and Images.
The reason behind this behavior could be the increased level of rich media inclusion for desktop clients (e.g., using Flash), whereas automatic adjustments performed (e.g., using Java Script) disable this for mobile clients, where this is not desired. Furthermore, we note that the average amount of bytes per image is fairly close (even a bit higher for mobile), which could be due to increased utilization of frameworks that change the page layouts on the fly based on clients, but not the actual contents (e.g., due to theming on popular content management systems), which results in identical downloads.

\subsection{Caching of Web Page Objects}
One of the main assumptions for the prior evaluation was that the data represents ``first views,'' i.e., the dataset's underlying measurements do not consider the caching of web pages after initial visits. 
The individual entries in the datasets obtained from httparchive.org, however, additionally contain the max-age header (amongst other, such as expiration) directives for evaluation of the maximum lifetime of objects on the requesting device. The locally cached data can have a positive impact on the required network access, especially for mobile devices.

\begin{figure}
	\centering
	\subfloat[Desktop clients]{\includegraphics[width=.45\textwidth]{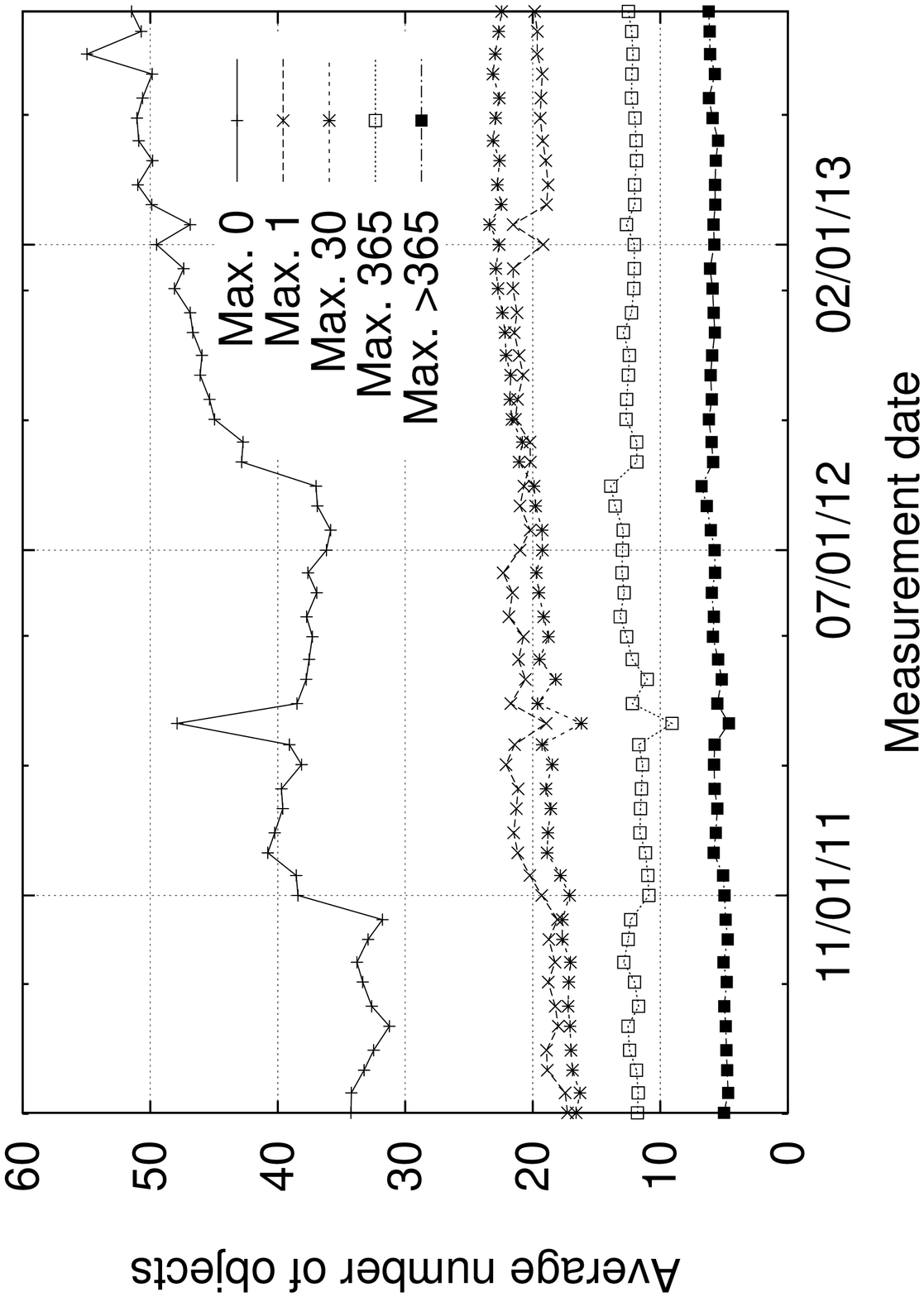}}\qquad
	\subfloat[Mobile clients]{\includegraphics[width=.45\textwidth]{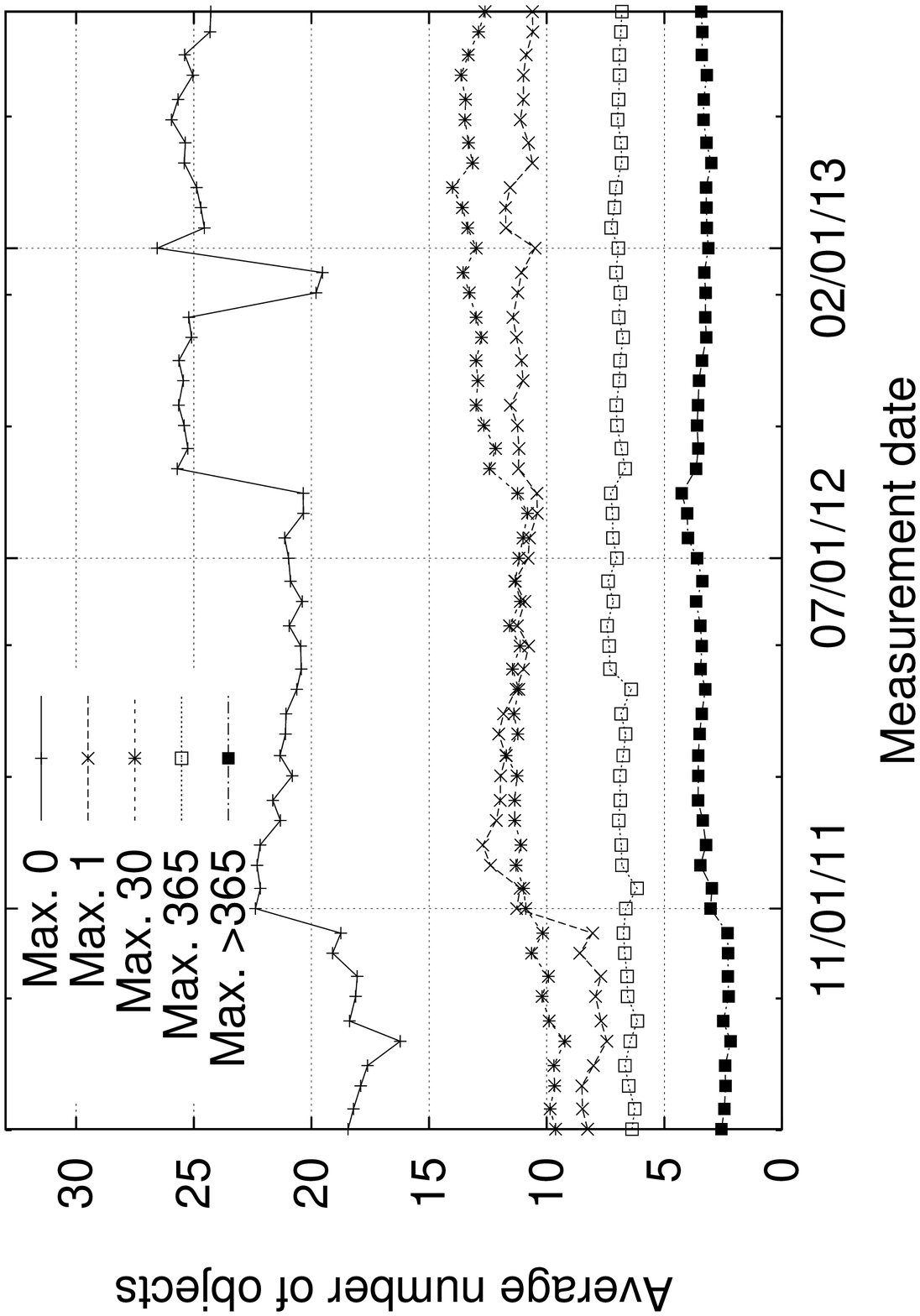}}\\
	\caption{Maximum age of web page objects for desktop and mobile web clients.\label{fig:maxage}}
\end{figure}

We illustrate the different max-age values obtained from the data in Figure~\ref{fig:maxage}~(a) for desktop client requests.
We observe that the largest fraction of items cannot be cached effectively, either as no caching is allowed or due to limits below a day, and in turn requires frequent downloads. 
More importantly, this is the only category that exhibits significant continued growth.
Over the time horizon we consider here, items that are stored between one and up to 30 days can be considered regularly downloaded at a comparable level, while only a relatively small number of objects can be considered invariant due to larger maximum allowable cache ages.
For the mobile client counterparts illustrated in Figure~\ref{fig:maxage}~(b), we initially note a similar level and continued growth for the average number of objects that cannot be cached (considering the overall average number of objects requested as discussed in Section~\ref{ss:objects}).
Objects that require regular, but less frequent downloads as well as more invariant items follow the trends observed for the desktop client.
Overall, the implication for mobile clients is that the fraction of requests with a shorter cache lifetime constitutes the majority of requests and in turn will require frequent downloads and in turn cause negative battery impacts.

\section{Implications}
\label{s:discuss}
One initial limitation of our dataset is the evaluation of landing pages only, which, however, could be seen as an actual upper limit, according to \cite{BuMaSe13}. 
Overall, we find significantly growing trends for the average web page number of objects and their sizes.
The immediate implications from the growing trend of the number of web request per fixed and mobile website versions and their respective size is that the pressure on access networks capacity growths will continue, see, e.g., \cite{Ci13} for a detailed separation into typical business and consumer categories of accesses Internet data.

We find significant similarities between the desktop and the mobile versions of the popular web pages that were part of this evaluation. 
Mobile counterparts of desktop page versions typically feature less objects and smaller sizes; both characteristics typically also exhibit a significantly slower growth.
As briefly outlined by the authors of \cite{BuMaSe13}, the reduced number of servers and number of objects involved likely can be seen as an indicator of the correlation between the two different versions of popular web pages.

\begin{figure}
	\centering
	\includegraphics[width=.475\linewidth]{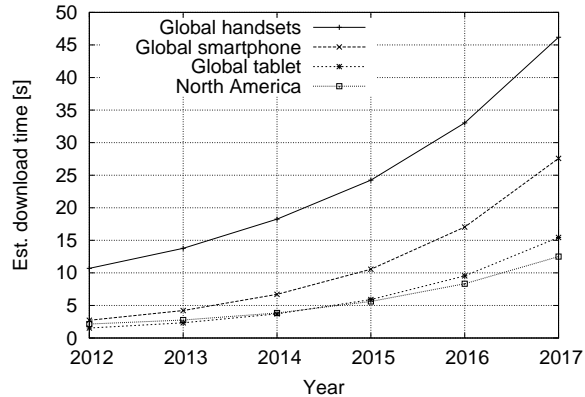}
	\caption{Approximation of future loading times for mobile web page clients.\label{fig:loading}}
\end{figure}

%
%
As previously outlined in~\cite{ThAgNiBoSi12}, Java Script objects in mobile web pages are a main contributor to the rendering power consumption, and we find significant amounts of Java Scripts to be contained in average web pages.
More importantly, we conceive that there is a significant amount of data and requests in average mobile pages that is dedicated to image and other objects, which will contribute significantly to the combined transmission and rendering energy costs, as outlined in \cite{ThAgNiBoSi12} as well.
Optimization efforts that target mobile clients specifically should in turn focus on these categories of data, specifically while taking more advantage of caching possibilities, as the transmission energy contributes a significant amount to the total energy required.
%
%
Combining the two prior evaluation categories, we noted that almost unforeseeable, the average number of bytes per request for desktop and mobile web pages are very similar and exhibit a linearly growing trend.
While some of the greater details highlight smaller differences by respective type of object requested, the overall direct correlation between the two gives rise to future research questions.
%
%
Lastly, we consider the long-term impact of the current growth in the average mobile web sizes with respect to loading times. 
Combining the forecast of average global network access speeds for all hand-held devices, smartphones, and tablets in \cite{Ci13} and the average web page size in Section~\ref{ss:bytes}, we illustrate a rough approximation of the average download time of mobile web pages in Figure~\ref{fig:loading}.
The wait time that average mobile users will have to endure will increase to about 45 seconds based on these projections. 
At the same time, there is little light for global smartphone users, who on average would be required to wait just short of 30 seconds.
Only global tablet users and Northern America mobile clients would experience waiting times around or below 15 seconds in this projection; according to \cite{NiUeNa10}, this amount of waiting time is already close to the bearable amount of time for users of web services.
Clearly, these forecasting combinations are not considering disruptive technology changes that might occur; they do, however, illustrate the need for more optimization and, likely, client considerations when designing web pages.

\section{Conclusion}
\label{s:conc}
This overview represents the first comparative study of the current state and changes of the World Wide Web, as seen by current fixed and mobile clients when requesting web pages.
We utilized the publicly available data from httparchive.org, an industry-supported web page statistics archival project, to outline major trends and resulting implications.
We performed and extensive analysis for web pages that were available in a desktop as well as a mobile version, allowing for direct comparisons of their average characteristics over time.
For the number of web-page requests, we noted an overall growing trend, with a slightly higher rate for desktop versions of web pages. Modern web desktop page versions feature well above 100 object requests on first views, while mobile versions trail in the range of 60.
Similarly, we noted a rapidly increasing trend with regard to the average web page sizes, with desktop pages approaching 1.5 MB in initial view size, while mobile pages trail at more than 50\%.
Most astonishingly, this study found that the average bytes per request are within close proximity for fixed and mobile clients and exhibit both comparable as well as significant growth rates.
When performing decomposition into categories, we found that Image, Other (such as Flash or fonts), and Java Script objects are the main contributors to web page sizes for both clients. Furthermore, the most requests for objects on first view are attained for the Image category of web page objects, again independent of the client type.
We also evaluated the potential of caching objects for subsequent views of web pages, whereby we found that most web page objects exhibit very short cache life times of a one day maximum.
Given recent mobile access network forecasts, we derive a drastic increase of waiting times, which could have significant impact on the user experience for mobile web access as well as a potential negative impact on mobile power consumption.
Our overview thus suggests that there are significant opportunities for future optimization efforts for efficient mobile web content delivery.


\section*{Acknowledgment}
We are grateful for insightful discussions with Prof. Martin Reisslein from Arizona State University.



\end{document}